\documentclass{article}
\usepackage[T1]{fontenc} 
\usepackage[utf8]{inputenc} 
\usepackage{ismir,amsmath,cite,url}
\usepackage{graphicx}
\usepackage{color}
\usepackage{multirow}
\usepackage{booktabs}

\usepackage{lineno}

\usepackage{makecell}

\usepackage{float}
\usepackage{stfloats}

\title{HPPNet: Modeling the Harmonic Structure and Pitch Invariance in Piano Transcription}

\multauthor
{Weixing Wei$^1$ \hspace{1cm} Peilin Li$^1$ \hspace{1cm} Yi Yu$^2$ \hspace{1cm} Wei Li$^1$$^3$} {
 $^1$ School of Computer Science and Technology, Fudan University , China\\
$^2$ Digital Content and Media Sciences Research Division, National Institute of Informatics (NII), Japan\\
$^3$  Shanghai Key Laboratory of Intelligent Information Processing, Fudan University, China\\
{\tt\small wxwei20@fudan.edu.cn, plli21@m.fudan.edu.cn, yiyu@nii.ac.jp, weili-fudan@fudan.edu.cn}
}

\def\authorname{Weixing Wei, Peilin Li, Yi Yu, and Wei Li}

\begin{document}

\maketitle
%
%

\begin{abstract}
While neural network models are making significant progress in piano transcription, they are becoming more resource-consuming due to requiring larger model size and more computing power. In this paper, we attempt to apply more prior about piano to reduce model size and improve the transcription performance.
The sound of a piano note contains various overtones, and the pitch of a key does not change over time. 
To make full use of such latent information, we propose HPPNet that using the Harmonic Dilated Convolution to capture the harmonic structures and the Frequency Grouped Recurrent Neural Network to model the pitch-invariance over time. Experimental results on the MAESTRO dataset show that our piano transcription system achieves state-of-the-art performance both in frame and note scores (frame F1 93.15\%, note F1 97.18\%). Moreover, the model size is much smaller than the previous state-of-the-art deep learning models. 
\end{abstract}

\section{Introduction}
\label{sec:intro}
Automatic music transcription (AMT) is a crucial task in Music Information Retrieval (MIR). This task converts the music in audio format to musical notation formats  such as MIDI and sheet music. Transcribing from wave format is a process of message compression that reduces the message from universe form (wave) to abstract form (sheet music) that will help with music understanding.  

Piano transcription is a popular subtask of AMT. Predicting a set of concurrent pitches present in the same frame is a challenging problem. 
Over the past decades, plenty of methods have been applied to the piano transcription task, e.g., using Factorization-based models (Smaragdis et al.\cite{non-negative}), using sparsity coding and unsupervised analysis (Abdallah et al. \cite{Abdallah}),  adaptive estimation of harmonic spectra (Vincent et al. \cite{VincentBB10}), and using SVM-HMM structure (Nam et al. \cite{NamNLS11} ). Some methods are motivated by the piano's acoustics features, such as the attack/decay \cite{attack_decay} model.

In recent years, with the development of deep learning and the existence of large scale labeled datasets, neural networks have become a popular method, such as using RNNs \cite{BockS12} and using methods based on CNNs \cite{SigtiaBD16}.
With the release of the MAESTRO \cite{maestro} dataset in the field of piano transcription, the Onsets \& Frames transcription system \cite{Onsets_Frames} made significant progress. Hawthorne et al. focus on onsets and offsets together to predict frame labels.  Transformer is a revolutionary architecture in other fields, and Hawthorne et al. \cite{transformer} explore Transformer’s potential on piano transcription. Generative Adversarial Networks \cite{Adversarial} also show its potential in improving performance. Kong et al. \cite{high_resolution} proposed a high-resolution AMT system trained by regressing precise onset and offset times of piano notes, which achieved state-of-the-art performance in note prediction.


\begin{figure}[t]
\centering
\includegraphics[scale=0.23]{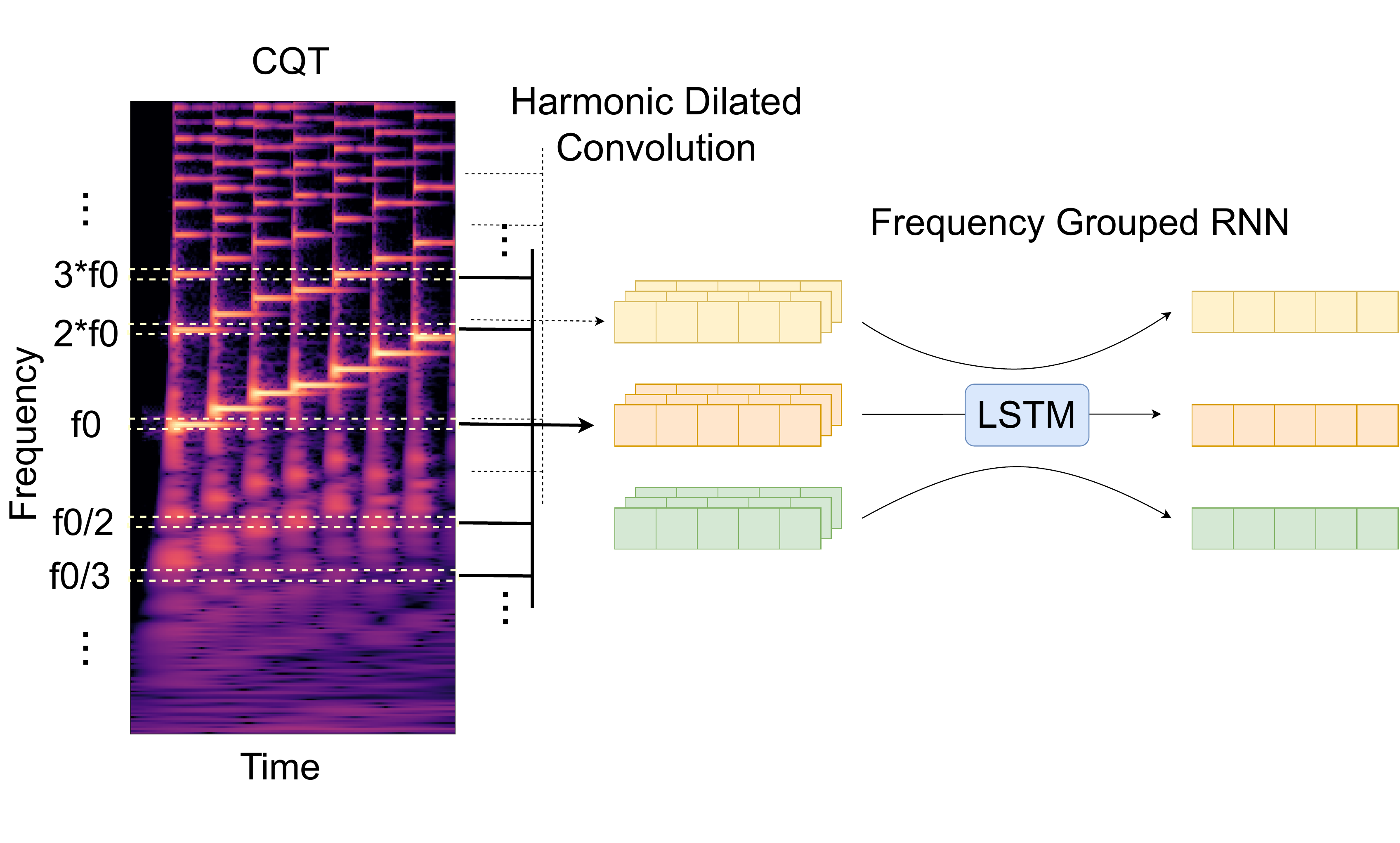}
\caption{Harmonic Dilated Convolution and Freqency Grouped RNN. The former captures possible harmonic series information for all frequency groups by dilated convolution. The latter feeds each single frequency group to the same LSTM layer to detect whether there is a harmonic series in the frequency group. }
\label{fig:harmonic_dilated_convolution}
\end{figure}

However, previous SOTA models are becoming more and more resource-consuming.
The spectrograms of the solo piano are highly structured. Each tone has a harmonic series, and the pitch of a key does not change over time. 
In view of this prior, we attempt to model such harmonic structure and pitch-invariance over time to improve model interpretability and algorithm efficiency.

We propose the Harmonic Dilated Convolution (HD-Conv) to capture harmonic series information and the Frequency Grouped Long Short-Term Memory (FG-LSTM) to detect if there is an active pitch accounting for high energy in a specific frequency band.
We apply HD-Conv and FG-LSTM to model the harmonic structure and pitch-invariance prior over time in a model called HPPNet. The model achieves state-of-the-art piano transcription performance. Remarkably, it is an approach that uses much fewer parameters of only 1.2 million while other models like the Transformer\cite{transformer} use 54 million ones. 
The reduction of parameters is attributed to the use of FG-LSTM, which applies LSTM in a single frequency group, and the shared acoustic model.
The primary contribution of this work is an tiny model that achieves state-of-the-art performance using much fewer parameters. 

The rest of this paper is organised as follows: Section \ref{sec:architecture} describes the proposed model component's details: harmonic dilated convolution and the frequency grouped recurrent neural networks. Section \ref{sec:experiments} illustrates the experimental setup, with results analysed in Section \ref{sec:result}. Conclusions are discussed in Section \ref{sec:conclusion}.

\section{Architecture}
\label{sec:architecture}

The two critical components of our model are the harmonic dilated convolution and the frequency grouped recurrent neural networks.  The former is used to model harmonics structure, and the latter is designed based on the invariance of piano pitches over time as demonstrated in Figure \ref{fig:harmonic_dilated_convolution}.

\subsection{CQT input}
We use the Constant-Q Transform (CQT) \cite{CQT} as our input feature. Unlike the log-mel spectrogram which is partially log-scaled, all the frequency bins of the CQT are geometrically spaced. As shown in Eq. (\ref{eq:log_intervals}), the spacing between fundamental frequency and overtones does not vary with the change of the fundamental frequency. Such property makes it suitable for dilated convolution, as described in \cite{HarmoF0}.

\begin{equation}
\begin{aligned}
d_{k} &= log_{2^{1/Q}} (k \cdot f_0) - log_{2^{1/Q}} (f_0) \\
&= Q \cdot log_{2}(k),
\end{aligned}
\label{eq:log_intervals}
\end{equation}
where $k$ is the serial number of the harmonic series, $d_{k}$ denotes the distance between fundamental frequency and the k-th overtone on a log frequency scale, $Q$ indicates the number of frequency bins per octave, and $f_0$ is the fundamental frequency.



\subsection{Harmonic Dilated Convolution}
\begin{figure}[htb]
\centering
\includegraphics[scale=0.47]{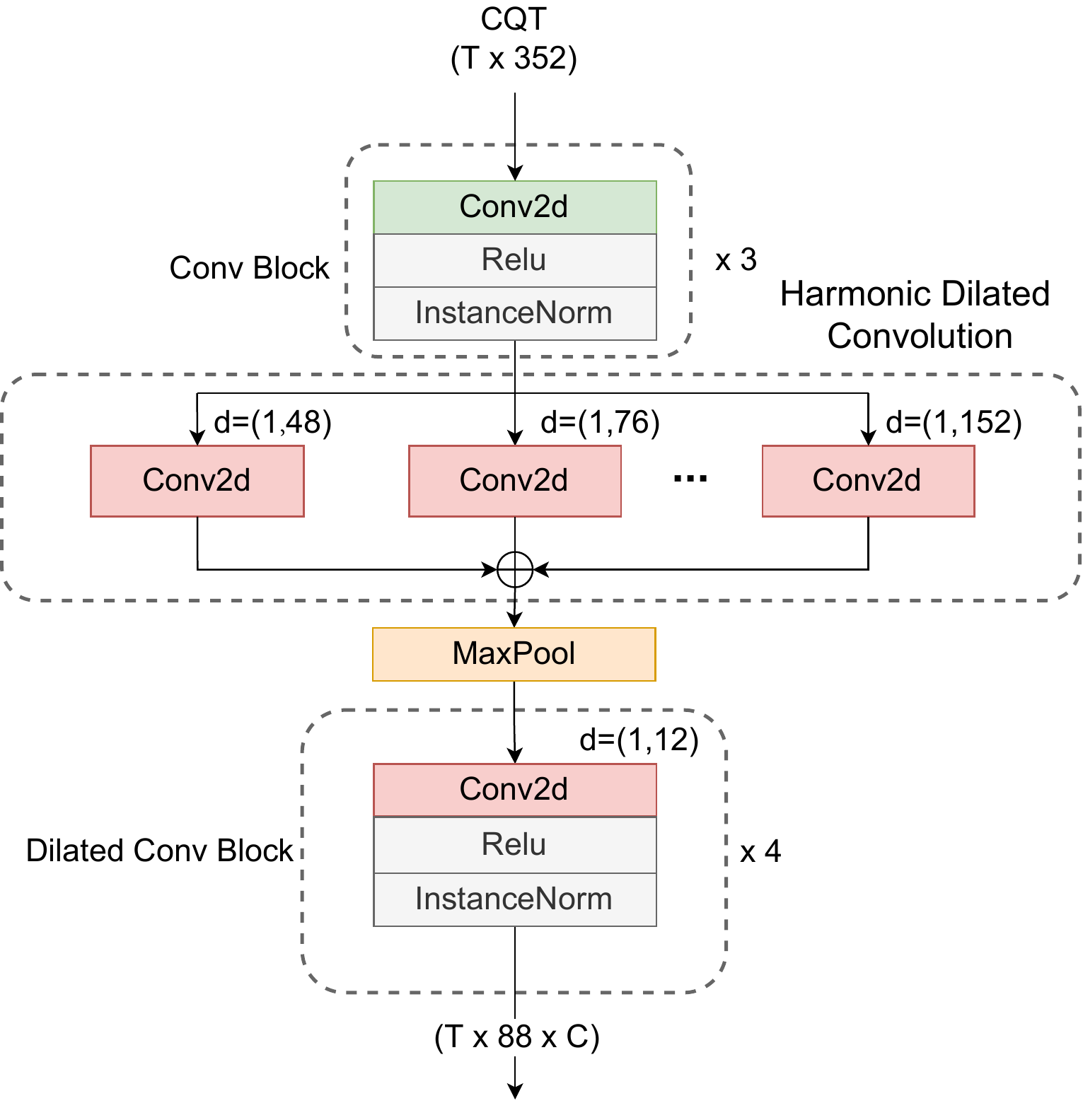}
\caption{Acoustic model of HPPNet. It is composed of three identical regular convolution layers, a harmonic dilated convolution layer, a max pooling layer, and four identical dilated Conv blocks. Different dilated convolution layers have different dilation rates denote by $d$. The output is a feature map with size $T \times 88 \times C$, where $T$ is the frame numbers, 88 denotes the numbers of piano keys, $C$ is the channel size. }
\label{fig:acoustic_model}
\end{figure}

In recent years, some methods are proposed to modeling harmonic structure in neural networks.
The Harmonic constant-Q transform (HCQT) \cite{HCQT} captures the harmonic relationships by a 3-dimensional CQT array for pitch tracking in polyphonic music. Harmonic convolution is applied in \cite{harmonic_convolution} to denoise speech audios. Dilated convolution \cite{WangLS20a} and sparse convolution \cite{WangLS20_harmonic} are used in Wang's works but they are still not perfect ways to capture harmonic structure. Our model captures harmonic information in a simple but effective way. As shown in our acoustic model (Figure \ref{fig:acoustic_model}), we feed the CQT into multiple dilated convolution \cite{dilated_convolution} layers with different dilation rates and sum the outputs for the following layers. The dilation rates are the distance between fundamental frequency and overtones on a log frequency scale as in Eq. (\ref{eq:log_intervals}). We call such dilated convolutions applied on the log-frequency dimension and with dilation rates of spaces between harmonic series the Harmonic Dilated Convolution (HD-Conv).


\subsection{Frequency Grouped LSTM}


In the feature map output by the harmonic acoustic model, each frequency unit with multiple channels contains corresponding possible harmonic information to detect if there is an active pitch accounting for high energy in a specific frequency band. 
However, detecting multiple pitches in polyphonic music is challenging since harmonic series interfere with each other.
Time-domain correlation is required to obtain smooth pitch contours.
Previous works \cite{Onsets_Frames}\cite{high_resolution} applied bidirectional Recurrent Neural Network (biRNN) \cite{biRNN} to model the temporal relationship of frames. They flatten the channel dimension and frequency dimension of the feature map output by the acoustic model to a long hidden dimension as the input of RNN. There are some problems with such processing. One is that the long hidden dimension led to an unnecessarily large amount of model parameters. This leads to large amount of parameters both in acoustic model and Long Short-Term Memory (LSTM) \cite{LSTM}, as described in Table \ref{tab:model_parameters}.


\begin{table}[t]
 \begin{center}
 \resizebox{.8\width}{!}{
 \renewcommand{\arraystretch}{1.3}
 \begin{tabular}{ c |c c c c c c}
 \toprule[2pt]
 \multirow{2}{*}{Model}  & \multicolumn{2}{c}{Onsets \& Frames} & \multicolumn{2}{c}{HPPNet}\\
 \cmidrule(r){2-3} \cmidrule(r){4-5}
 
 & Acoustic & LSTM & Acoustic & FG-LSTM  \\
 
\hline

inputs & T$\times$229 & T$\times$768 & T$\times$352 & T$\times$128  \\
outputs & T$\times$768 & T$\times$88 & T$\times$88$\times$128 & T$\times$1\\
units & - & 256 & - & 128\\
\hline
parameters & 4.3M & 3.5M & 421K & 99K \\


 
 \bottomrule[2pt]
 \end{tabular}
 }
\end{center}
 \caption{The parameters of different layers in Onsets \& Frames and HPPNet. Acoustic denotes the acoustic models, T denotes the number of frames in a training sample.} \label{tab:model_parameters}
\end{table}


The sub-band and multi-band techniques are widely applied in speech and music processing \cite{multi_scale_multi_band, multi_channel_speech, performance_net}. The common practice is splitting the full-band spectral representation into multiple sub-bands and processing them separately, then concatenating the outputs of each sub-band for subsequent processing.  Specifically, some methods based on splitting sub-bands (frequency-LSTM \cite{group_comm}) or  generating sub-bands (multi-band MelGAN \cite{multi_band_melgan}) show impressive performance with lightweight models. The splitting of sub-bands reduces the size of network input, and different sub-bands sharing the same sub-network further reduces the model size. 

Similar to the multi-bands techniques, we segment the output of harmonic dilated convolution to 88 frequency groups.
Since the piano is a fixed-pitch keyboard instrument, the fundamental frequency of each key does not vary as time changes.
This enables the model process each frequency group independently.
Based on this assumption,
we feed each frequency group to the same LSTM layer individually to model the temporal relationship. We term this the Frequency Grouped LSTM (FG-LSTM) illustrated in Figure \ref{fig:harmonic_dilated_convolution}. Feeding a single frequency group to the LSTM makes inputs and units size of FG-LSTM smaller. This method significantly reduces the amount of parameters in our model.

\subsection{Model Details}

Raw audios are clipped to 20-second pieces and resampled to 16 kHz. Then, the CQT is computed by nnAudio \cite{nnAudio} with hop length 320 (20 millisecond), an FFT window of 2048, bins per octave of 48, fmin of 27.5 Hz, frequency bins number of 352, and log amplitude.
The model takes the CQT feature with size $T \times 352 $ as input, where T is the frame number of each audio clip. The time resolution and frequency resolution of inputs are limited by the computational resource, higher input resolution may have better performance.

The model is composed of a convolutional acoustic model and multiple FG-LSTM heads. The former structure captures the harmonic representations of a tone, and the latter establishes connections over temporal series.
In the acoustic model shown in Figure  \ref{fig:acoustic_model}, the first three convolution layers extract local information with kernel size $ 7\times 7 $, a ReLu activation, and instance normalization.

Then followed by eight dilated convolution layers  parallel with different dilations of 48, 76, 96, 111, 124, 135, 144, and 152  in the frequency dimension. These dilation rates are calculated by Eq.  (\ref{eq:log_intervals}) with $Q=48$ and rounded to integers. All the outputs of these layers integrate the harmonic information for each frequency unit. 

The max-pooling layer downsamples the frequency bins from 352 to 88 with pooling size of 4, four bins per semitone to one bin per semitone. Then four identical dilated convolution blocks make the network deeper to achieve higher learning capacity. Each block contains one convolution layer with kernel size of $ 5 \times 3 $, and dilation of $ 1 \times 12 $.


\begin{figure}[htb]
\centering
\includegraphics[scale=0.47]{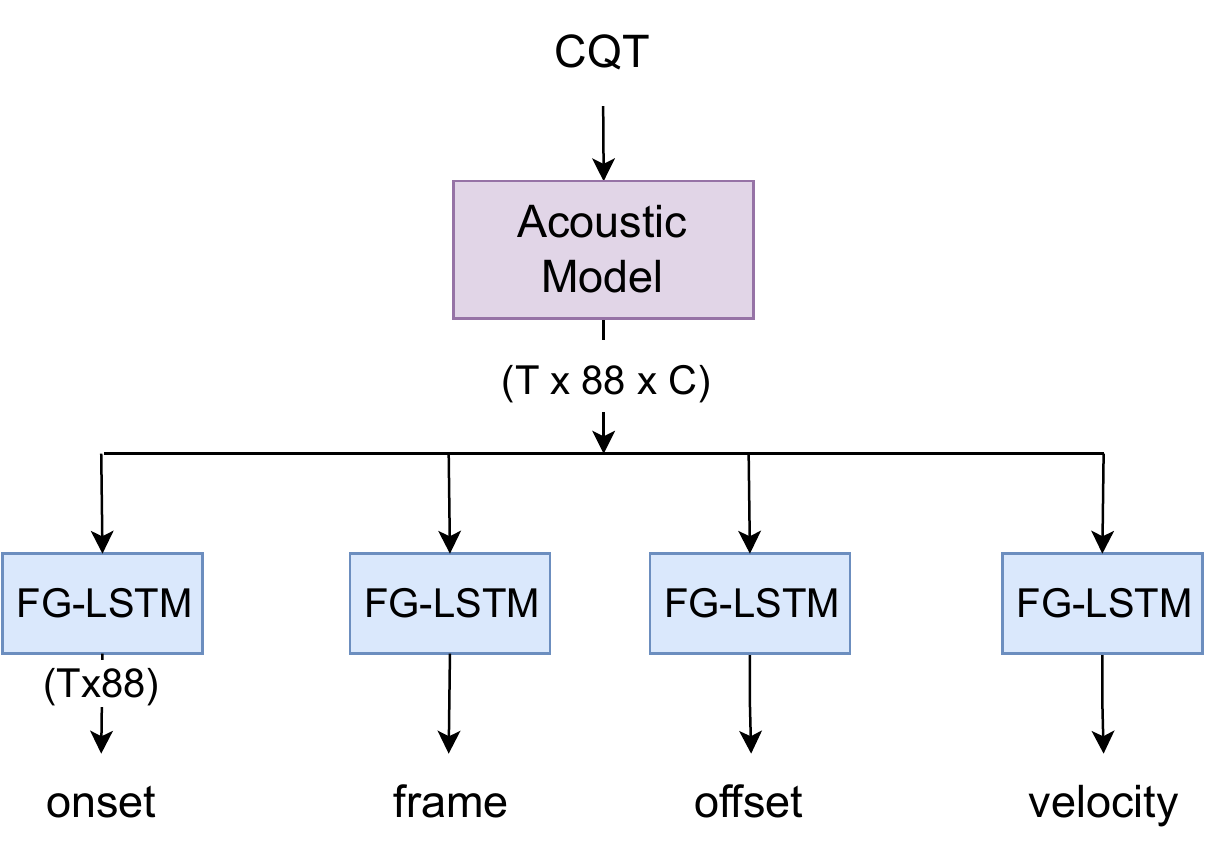}
\caption{HPPNet-base. Onset, frame, offset, and velocity tasks share the same acoustic model.}
\label{fig:HPPNet_base}
\end{figure}

Finally, four FG-LSTM heads are used to predict frames, onsets, offsets, and velocities separately as shown in Figure \ref{fig:HPPNet_base}. These outputs are then decoding to note events with the same way as the Onsets \& Frames \cite{maestro} model. Note that the offset head is just used for training and not directly used during decoding. Each head uses a linear layer with a sigmoid activation function to output the prediction. All the outputs are matrices of size $ T\times 88 $, where $T$ denotes frame number and 88 represents the 88 piano keys. The LSTM is bidirectional, and both forward and backward directions have 64 units. The hyperparameters of the network are summarized in Table \ref{tab:hyperparameters}.
 We denote this model the HPPNet-base.


\begin{table}[htb]
 \begin{center}
  \resizebox{.85\width}{!}{
\begin{tabular}{c m{1.6em} c m{1.6em} c}
\toprule[1.5 pt]
 Layer          & repeat & kernel & filters & dilation\\
 \hline
Conv           & 3      & 7$\times$7    & 16          & -    \\
\hline
HD-Conv        & 1      & 1$\times$3    & 128         &\makecell{ 48, 76, ..., 152} \\
\hline
Dilated Conv & 4      & 5$\times$3    & 128         & 1$\times$12  \\
\hline
\end{tabular}

}
\end{center}
\caption{Hyperparameters of convolution layers.}
\label{tab:hyperparameters}
\end{table}

\begin{figure}[htb]
\centering
\includegraphics[scale=0.47]{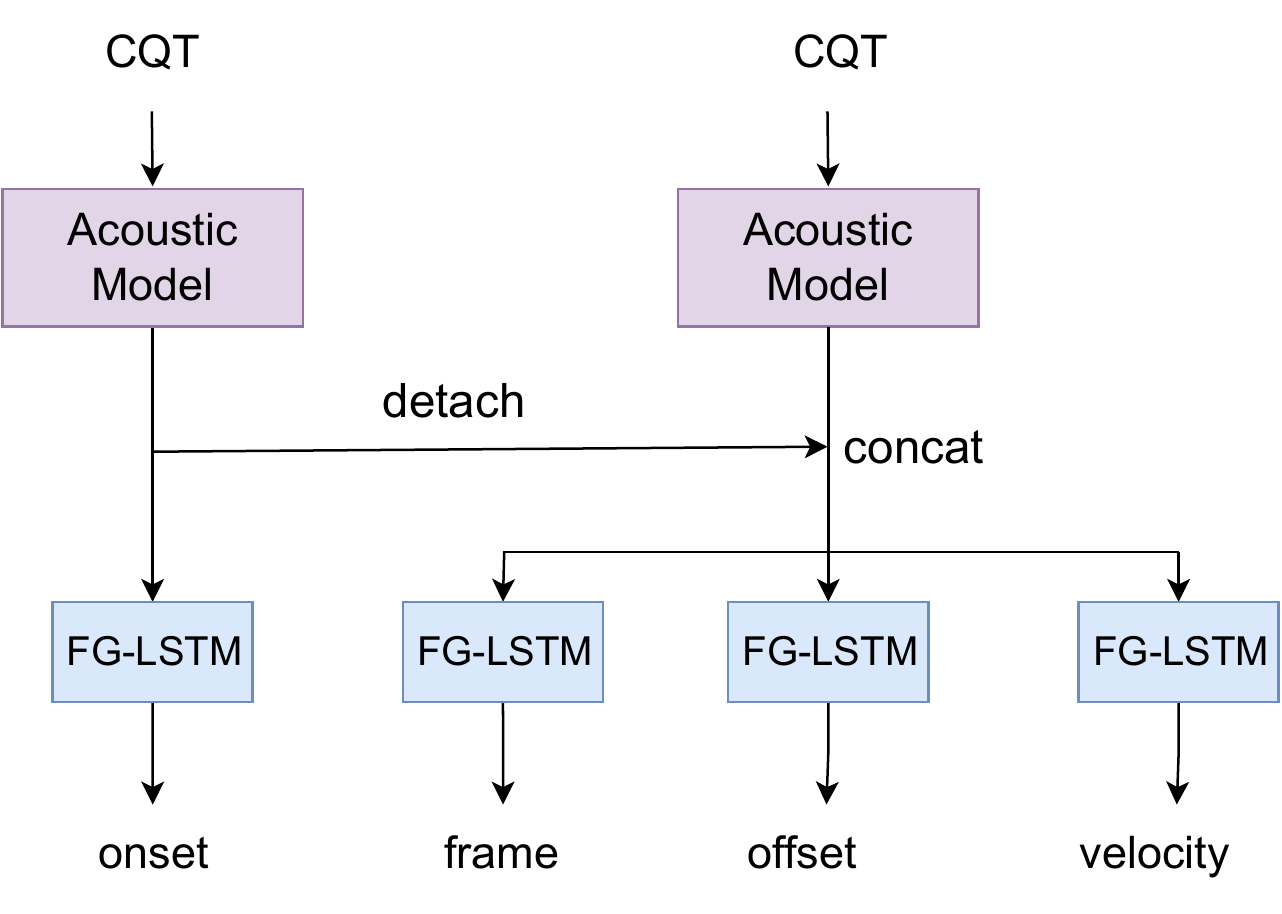}

\caption{HPPNet-sp. Onset and other tasks use separate acoustic models.}
\label{fig:HPPNet_sp}
\end{figure}
%
All the FG-LSTM heads in HPPNet-base share the same acoustic model, making it simple and efficient. But in our experiments, we try multiple acoustic models and find that if a stand-alone acoustic model is used to predict the onset, the model will perform better in onset detection. Since the onset F-measure is the metric that correlates best with human judgment \cite{YcartLBP20}, we use an individual acoustic model for onset. Frame, offset, and velocity share another acoustic model. The HPPNet-sp is shown in Figure \ref{fig:HPPNet_sp}. The output representations of the onset acoustic model are concatenated with outputs of the other acoustic model to predict frame, offset, and velocity (detach to avoid backward propagation).

\section{Experiments}
\label{sec:experiments}
In this section we first introduce the datasets we used. And then, we describe  the details of the experiment and the training loss for our model.

\subsection{Datasets}
MAESTRO \cite{maestro} (MIDI and Audio Edited for Synchronous Tracks and Organization): It includes about 200 hours of paired audio and MIDI recordings from ten years of International Piano-e-Competition. Audio and MIDI files are aligned with 3 ms accuracy and sliced to individual musical pieces annotated with composer, title, and year of performance. The MIDI have been collected by Yamaha Disklaviers which have a high-precision MIDI capture system.

MAPS \cite{MAPS} (MIDI Aligned Piano Sounds): It includes about 31 GB of CD-quality recordings and corresponding annotations of isolated notes, chords, and complete piano pieces. The records contain both synthesized audio and real audio by Virtual Piano software and a Yamaha Disklavier respectively.

\subsection{Experimental setup}

We use PyTorch \footnote{www.pytorch.org} to implement our model. The training is performed by Adam \cite{Adam} optimizer with mini-batch size 4, and a learning rate of $0.0006$ for 200k to 500k steps with early stopping. The training time on an NVIDIA GeForce 3060 GPU with 12 GB VRAM is about 48 hours. The note and frame scores are calculated by the mir\_eval library \cite{mir_eval}. The evaluation uses a 50 ms tolerance for note onset and an offset ratio of 0.2 as the default configuration in mir\_eval. The onset and frame thresholds are both set to 0.4 on the sigmoid output in our model. All the hyperparameters are selected by the performance on the validation split of MAESTRO.

The losses we use for training are similar to the Onsets \& Frames model. The losses of onset, frame, and offset are binary cross-entropy losses. 
The weighted frame loss is not used in our training. The loss of velocity is the mean squared error loss. All losses are summed together as the final loss. All the losses are as follows:

\begin{equation}
\begin{aligned}
l_{bce}\left(y \ , \  \hat{y}\right) =-wy \log (\hat{y})-(1-y) \log (1-\hat{y}),
\end{aligned}
\label{eq:onset_loss}
\end{equation}


\begin{equation}
\begin{aligned}
L_{onset}=\sum_{p=1}^{88} \sum_{t=1}^{T} l_{bce}\left(n_{p, t} \ , \  \hat{n}_{p, t}\right),
\end{aligned}
\label{eq:onset_loss}
\end{equation}

\begin{equation}
\begin{aligned}
L_{frame}=\sum_{p=1}^{88} \sum_{t=1}^{T} l_{bce}\left(f_{p, t} \ , \  \hat{f}_{p, t}\right),
\end{aligned}
\label{eq:frame_loss}
\end{equation}

\begin{equation}
\begin{aligned}
L_{offset}=\sum_{p=1}^{88} \sum_{t=1}^{T} l_{bce}\left(o_{p, t} \ , \  \hat{o}_{p, t}\right),
\end{aligned}
\label{eq:offset_loss}
\end{equation}




\begin{equation}
\begin{aligned}
L_{velocity}=\sum_{p=1}^{88} \sum_{t=1}^{T} n_{p, t}\left(v_{p,t} - \hat{v}_{p, t}\right)^{2},
\end{aligned}
\label{eq:vel_loss}
\end{equation}

\begin{equation}
\begin{aligned}
L=L_{onset} + L_{frame} + L_{offset} + L_{velocity}.
\end{aligned}
\label{eq:total_loss}
\end{equation}
where $l_{bce}$ denotes the binary cross entropy loss, $p$ is the piano note index range from 1 to 88, $T$ is the frame number in a sample, $w$ denotes the positive weight ($w=2$ for onset, $w=1$ for frame and offset), $y$ denotes ground true and  $\hat{y}$ \ denotes predicted values range from 0 to 1, $n_{p,t}$, $f_{p,t}$, $o_{p,t}$, and $v_{p,t}$ are the ground true of onset, frame, offset, and velocity seperately.


\subsection{Baselines}

Our model is compared with five typical models, including High-Resolution piano transcription \cite{high_resolution}, Onsets \& Frames \cite{Onsets_Frames},  Adversarial onsets \& frames \cite{Adversarial}, Semi-CRFs \cite{skip_frame} and the sequence-to-sequence Transformer \cite{transformer}. The results are shown in Table \ref{tab:f1}. High-Resolution and Onsets \& Frames are convolution-based models composed of convolution layers and LSMT layers. Especially,  the High-Resolution is the best model in note F1 score which is 96.72\%. Adversarial onsets \& frames is an adversarial training scheme that operates on the Mel-spectrogram. Semi-CRFs is a Semi-Markov Conditional Random Fields (semi-CRF) based model, which treats note intervals as holistic events. Sequence-to-sequence Transformer uses a generic Transformer architecture with standard decoding methods.

\begin{table*}[htb]
 \begin{center}
 \resizebox{.75\width}{!}{
 \renewcommand{\arraystretch}{1.3}
 \begin{tabular}{c c |c c c c c c c c c c c c}
 \toprule[2pt]
 \multirow{2}{*}{Model} & \multirow{2}{*}{Params} & \multicolumn{3}{c}{FRAME} & \multicolumn{3}{c}{NOTE} &
 \multicolumn{3}{c}{NOTE W/OFFSET} & \multicolumn{3}{c}{NOTE W/OFFSET \& VEL.} \\
 
 \cmidrule(r){3-5} \cmidrule(r){6-8} \cmidrule(r){9-11} \cmidrule(r){12-14}
 
 & & P (\%) & R (\%) & F1(\%) & P (\%) & R (\%) & F1(\%) 
 & P (\%) & R (\%) & F1(\%) & P (\%) & R (\%) & F1(\%) \\
 
\midrule[1.5pt]
 && \multicolumn{12}{c}{MAESTRO v1}\\
 
\hline
Onsets \& Frames \cite{maestro} & 26M & 92.11 & 88.41 & 90.15 & 98.27 & 92.61 & 95.32 &
82.95 & 78.24 & 80.50 & 79.89 & 75.37 & 77.54 \\

\hline
Adversarial onsets  \& frames \cite{Adversarial} & 26M & 93.1 & 89.8 & 91.4 & 98.1 & 93.2 & 95.6 &
83.5 & 79.3 & 81.3 & 82.3 & 78.2 & 80.2 \\

\midrule[1.5pt]
 && \multicolumn{12}{c}{MAESTRO v2}\\

\hline
High-Resolution \cite{high_resolution} & 20M & 88.71 & 90.73 & 89.62 & 98.17 & 95.35 & 96.72 & 
83.68 & 81.32 & 82.47 & 82.10 & 79.80 & 80.92 \\

\hline
Semi-CRFs \cite{skip_frame} & 9M & 93.85 & 88.72 & 91.11 & 98.66 & 94.50 & 96.51 & 90.68 & 86.89 & \pmb{88.72} & 89.68 & 85.96 & \pmb{87.75}\\


\hline
HPPNet-sp & 1.2M & 92.36 & 93.46 & \pmb{92.86} & 98.31 & 96.18 & \pmb{97.21} & 85.36 & 83.54 & 84.41 & 83.85 & 82.08 & 82.93 \\

\midrule[1.5pt]
 && \multicolumn{12}{c}{MAESTRO v3}\\
 
 \hline
Onsets \& Frames [reproduced] & 26M & 94.43 & 85.50 & 89.68 & 98.67 & 92.08 & 95.22 &
82.25 & 76.88 & 79.44 & 80.80 & 75.55 & 78.05 \\

\hline
Transformer \cite{transformer} & 54M & - & - & - & - & - & 96.13 &
- & - & 83.94 & - & - & 82.75 \\

\hline
Semi-CRFs \cite{skip_frame} & 9M & 93.79 & 88.36 & 90.75 & 98.69 & 93.96 & 96.11 & 
90.79 & 86.46 & \pmb{88.42} & 89.78 & 85.51 & \pmb{87.44}\\

\hline
HPPNet-tiny & 151K & 92.26 & 91.98 & 92.06 & 96.75 & 94.94 & 95.82 &
83.77 & 82.26 & 83.00 & 82.77 & 81.29 & 82.01 \\
\hline
HPPNet-base & 820K & 92.35 & 92.75 & 92.51 & 97.60 & 94.90 & 96.25 & 
84.03 & 83.48 & 83.57 & 82.94 & 81.23 & 82.12 \\

\hline
HPPNet-sp & 1.2M & 92.79 & 93.59 & \pmb{93.15} & 98.45 & 95.95 & \pmb{97.18} & 84.88 & 82.76 & 83.80 & 83.29 & 81.24 & 82.24 \\
 
 \bottomrule[2pt]
 \end{tabular}
 }
\end{center}
 \caption{Transcription result evaluated on the MAESTRO dataset.  Train split of MAESTRO v3 is used for training and test splits of different versions are used for evaluation.}
 \label{tab:f1}
\end{table*}

\begin{table*}[htbp]
 \begin{center}
 \resizebox{.83\width}{!}{
 \renewcommand{\arraystretch}{1.3}
 \begin{tabular}{c c |c c c c c c c c c c c c}
 \toprule[2pt]
 \multirow{2}{*}{Model} & \multirow{2}{*}{Params} & \multicolumn{3}{c}{FRAME} & \multicolumn{3}{c}{NOTE} &
 \multicolumn{3}{c}{NOTE W/OFFSET} & \multicolumn{3}{c}{NOTE W/OFFSET \& VEL.} \\
 
 \cmidrule(r){3-5} \cmidrule(r){6-8} \cmidrule(r){9-11} \cmidrule(r){12-14}
 
 & & P (\%) & R (\%) & F1(\%) & P (\%) & R (\%) & F1(\%) 
 & P (\%) & R (\%) & F1(\%) & P (\%) & R (\%) & F1(\%) \\
 
\hline
Onsets \& Frames & 26M & - & - & 82.02 & - & - &  83.04 & - & - & 61.84 & - & - & 48.07 \\
 
\hline
Onsets \& Frames* & 26M & 92.86 & 78.46 &	84.91 &	87.46 &	85.58 &	86.44 &	68.22 &	66.75 &	67.43 &	52.41 &	51.22 &	51.77\\

\midrule[1.5pt]
HPPNet-sp & 1.2M & 88.42 & 86.81 & 87.56 & 91.61 & 82.38 & 86.63 & 65.01 & 63.84 & 64.39 & 60.35 & 59.26 & 59.77 \\
 
 \bottomrule[2pt]
 \end{tabular}
 }
\end{center}
 \caption{The transcription evaluation result on MAPS test split, which training was done on the MAESTRO v3 train split. Onsets \& Frames* means Onsets \& Frames with audio augmentation. The training of HPPNet is without audio augmentation.} \label{tab:maps}
\end{table*}

\section{Results}
\label{sec:result}

We compare our model with some existing state-of-the-art methods using the MAESTRO dataset and the MAPS dataset. 
Then, ablation studies are done to demonstrate the affects of FG-LSTM, and HD-Conv. We also examine the model performance on small datasets.

\subsection{Comparison with baselines}


Along with the HPPNet-base and HPPNet-sp, we also evaluate the HPPNet-tiny,  which has a similar structure as HPPNet-base but reduces the maximum convolution channel size and units in LSTM from 128 to 48.

Evaluation on the MAESTRO dataset is shown in Table \ref{tab:f1}. The HPPNet-sp achieves state-of-the-art in both frame F1 score and note F1 score, which improved 3.24 percentage points and 0.49 percentage points, reaching 92.86\% and 97.21\% in MAESTRO v2 respectively. In the last two columns, under the condition of note with offset, it still outperforms other models except the event-based Semi-CRFs method.
Even slight structure of HPPNet-base still achieves 92.51\% and 96.25\% on frame F1 score and note F1 score, respectively. 

Moreover, another significant advantage is that our model has much fewer parameters, around 820 thousand HPPNet-base and 1.2 million HPPNet-sp.
The lightweight structure HPPNet-tiny with 151 thousand parameters still outperforms the baseline Onsets \& Frames, that the frame F1 score and the note F1 score reaching 92.06\% and  95.82\%. 
This proves the effectiveness of HPPNet in both improving performance and reducing model size.

Empirical results in Table \ref{tab:f1} show that among all compared frame-level models, our model performs best in all scores. It is interesting to note that HPPNet achieves improved recall while not or only slightly loosing precision. But the event-based Semi-CRFs has better prediction on offsets, for it predicts note intervals directly rather than individual frames. Maybe future works that combining the frame-level and event-level can take advantage of the both sides.




\begin{figure}[h]
\includegraphics[width=7cm]{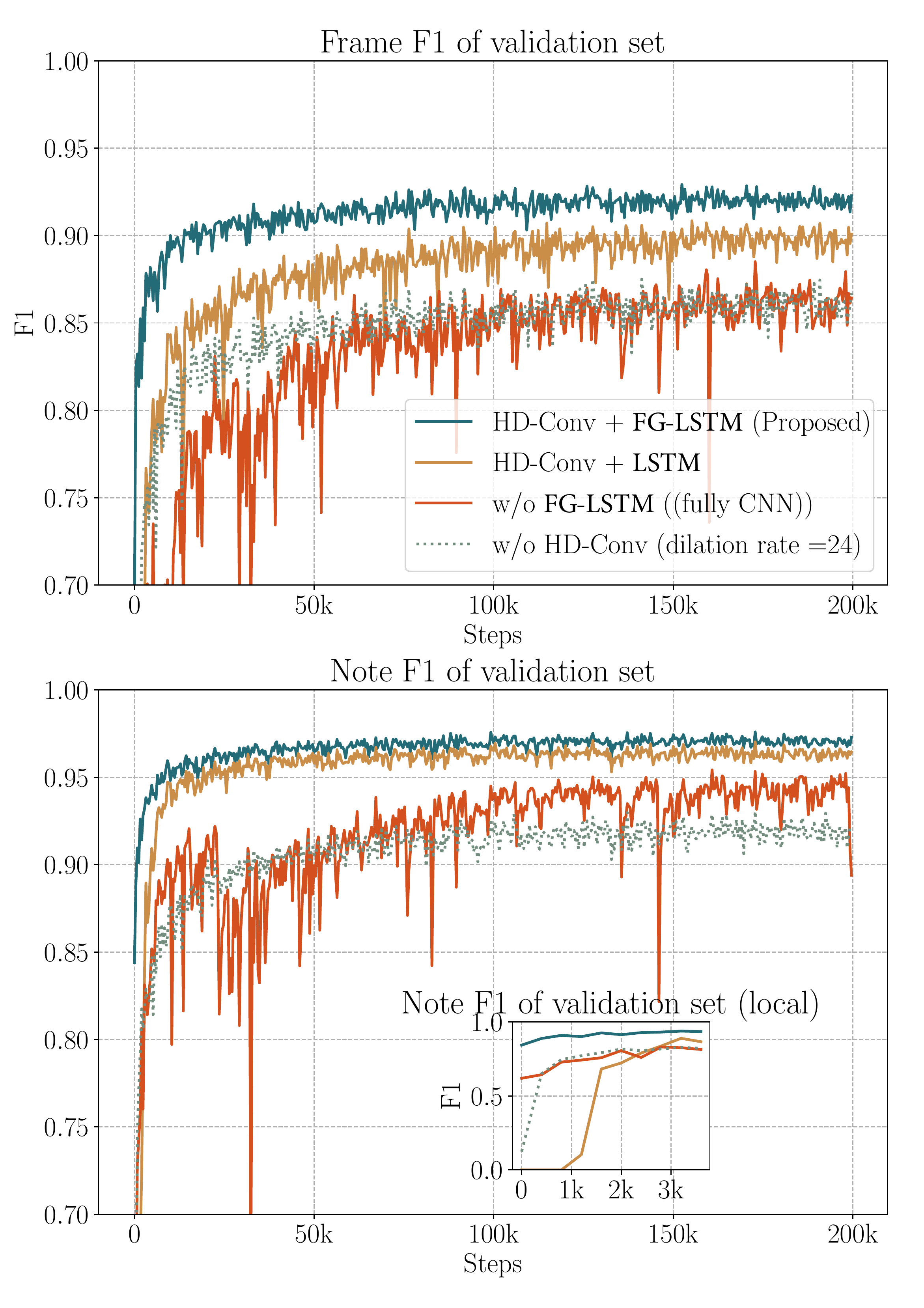}
\caption{F1 in frame-wise and note-wise of MAESTRO v3 validation set.}
\label{fig:ablation_train}
\end{figure}


To further explore the generalization of the HPPNet in different datasets, we train the model on the MAESTRO dataset and evaluate it on the MAPS dataset. The result is shown in Table \ref{tab:maps}. we can find that the HPPNet-sp's note F1 still reaches 87.56\%. This result is better than Onsets \& Frames of 83.04\%, even outperforming Onsets \& Frames with audio augmentation of 86.44\%. The result confirms that the HPPNet has better generalization ability on piano transcription task.






\subsection{Ablation study}
We further analyze the role of the HD-Conv layer and FG-LSTM in the HPPNet.
In addition to the HPPNet, we evaluate three types of models, the F1 in frame-wise and note-wise during training per epoch shown in Figure \ref{fig:ablation_train}:\\
\begin{itemize}
\item[$\bullet$] To verify the effect of FG-LSTM: change FG-LSTM to normal LSTM with frequency flattening. 
\item[$\bullet$] To verify the effect of LSTM: remove time-series layers, which change them to linear layers.
\item[$\bullet$] To verify the effect of harmonic dilated convolution: replace the harmonic dilated convolution with fixed dilated convolution (dilation rate of 24). 
\end{itemize}

These results suggest that the the HPPNet outperforms the other three ablated models. The F1 of normal LSTM decreases about three percentage points and one percentage point on frame-wise and note-wise, respectively. 
Usual LSTM starts later than the HPPNet in note-wise F1 at the beginning of training, and it starts to increase after about 1k steps.  
The HD-Conv + vanilla LSTM is better than changing time-series layers to linear layers. Model without RNN converges slowly and fluctuates drastically.
A large receptive field helps model get global information. Using the fixed dilation rate has a similar receptive field to HD-Conv. But the result shows it does not capture useful information for the detection of frames and onsets, leading to poor performance.



\begin{table}[!h]
 \begin{center}
 \resizebox{.73\width}{!}{
 \renewcommand{\arraystretch}{1.3}
 \begin{tabular}{c c |c c c c c c}
 \toprule[2pt]
 \multirow{2}{*}{Model} & \multirow{2}{*}{Data} & \multicolumn{3}{c}{FRAME} & \multicolumn{3}{c}{NOTE}\\
 \cmidrule(r){3-5} \cmidrule(r){6-8}
 
&& P (\%) & R (\%) & F1(\%) & P (\%) & R (\%) & F1(\%)  \\
 
\hline

\multirow{3}{1.7cm}{\centering Onsets\&\\Frames} 
& 100\% & 94.43 & 85.50 & 89.68 & 98.67 & 92.08 & 95.22 \\
& 30\% & 91.14 & 80.00 & 85.08 & 98.13 & 89.57 & 93.60\\
& 10\% & 90.66 & 72.70 & 80.36 & 96.64 & 85.46 & 90.62 \\

\hline
\multirow{3}{*}{HPPNet-sp }
& 100\% & 92.79 & 93.59 & \pmb{93.15} & 98.45 & 95.95 & \pmb{97.18} \\
& 30\% & 90.72 & 92.11 & 91.35 & 96.46 & 94.46 & 95.43\\
& 10\%  & 92.10 & 84.96 & 88.31 & 96.59 & 90.94 & 93.52\\
 
 \bottomrule[2pt]
 \end{tabular}
 }
\end{center}
 \caption{The transcription evaluation result on MAESTRO v3 test split, which training was done on the 100\%, 30\%, and 10\% of MAESTRO v3 training split respectively. Considering the annotation accuracy, we use subsets of MAESTRO rather than the MAPS for evaluation.} \label{tab:small_datasets}
\end{table}


\subsection{Performance on small datasets}

The HPPNet trained on big dataset such as compete MAESTRO shows promising results, and it also shows better generalization when inference on MAPS. To evaluate the performance of the HPPNet on smaller dataset, we also train the model with 30\% and 10\% of MAESTRO training split. All the samples are randomly selected and without data augmentations. The results are displayed in Table \ref{tab:small_datasets}. When the training set size decreases to 10\%, the HPPNet-sp drops less than Onsets \&  Frames on frame F1 and note F1 scores with 88.31\% and 93.52\% respectively, while the Onsets \& Frames approach only yields 80.36\% and 90.62\%. This demonstrates that the HPPNet relies less on data thanks to the small amount of parameters and priors contained in HD-Conv and FG-LSTM.

\section{Conclusion}
\label{sec:conclusion}

In this paper, we design a model that carries piano inductive biases to capture piano priors. The HPPNet is proposed to model the harmonic structure and the pitch-invariance over time in piano transcription. Experimental results show that the model achieves state-of-the-art performance on the MAESTRO dataset, with frame F1 of 93.15\% and note F1 of 97.18\%.  The success stems from two aspects: (i) the harmonic dilated convolution exploited to capture harmonics structure; and (ii) the frequency grouped LSTM designed based on the pitch-invariance of piano key over time. 
Furthermore, the model parameters are much fewer than the previous state-of-the-art models.

The results of the model on piano transcription are encouraging. And it may also be suitable for other instruments with similar pitch characteristics. But some challenges remain, such as improving performance in offset detection and modifying the model to other polyphonic transcriptions involving vocals and instruments with a less stable pitch. The harmonic dilated convolution implemented by multiple parallel dilated convolutions is computational consuming and needs to be optimized in the future.

\section{Acknoledgement}
This work was supported by National Key R\&D Program of China(2019YFC1711800), NSFC(62171138).  Wei Li and Yi Yu are corresponding authors of this paper.




\bibliography{ISMIRtemplate}

%
%
%
%
%

\end{document}